\documentclass[%
 reprint,
 amsmath,amssymb,
 aps,prl
]{revtex4-1}

\usepackage{graphicx}
\usepackage{epsfig}
\usepackage{dcolumn}
\usepackage{bm}
\usepackage{newfloat}
\DeclareFloatingEnvironment[name={S}]{suppfig}

\begin{document}


\title{Quantum Control by Imaging : The Zeno effect in an ultracold lattice gas}
\author{Y. S. Patil, S. Chakram and M. Vengalattore}
 \affiliation{Laboratory of Atomic and Solid State Physics, Cornell University, Ithaca, NY 14853}
\email{mukundv@cornell.edu}

\date{\today}

\begin{abstract}
We demonstrate the control of quantum tunneling in an ultracold lattice gas by the measurement backaction imposed by an imaging process. A {\em in situ} imaging technique is used to acquire repeated  images of an ultracold gas confined in a shallow optical lattice. The backaction induced by these position measurements modifies the coherent quantum tunneling of atoms within the lattice. By smoothly varying the rate at which spatial information is extracted from the atomic ensemble, we observe the continuous crossover from the `weak measurement regime' where position measurements have little influence on the tunneling dynamics, to the `strong measurement regime' where measurement-induced localization causes a large suppression of tunneling. This suppression of coherent tunneling is a manifestation of the Quantum Zeno effect. Our study realizes an experimental demonstration of the paradigmatic Heisenberg microscope in a lattice gas and sheds light on the implications of quantum measurement on the coherent evolution of a mesoscopic quantum system. In addition, this demonstrates a powerful technique for the control of an interacting many-body quantum system via spatially resolved measurement backaction. 
\end{abstract}

\maketitle

A fundamental distinction between a classical and a quantum system is its response to a measurement. While a classical system can be measured to arbitrary precision with negligible concomitant backaction, the act of measurement has profound consequences on the subsequent evolution of a quantum system \cite{braginsky99}. In the extreme limit, a sequence of rapid,  projective measurements can freeze the decay of an unstable quantum system \cite{fischer2001,syassen2008,yan2013}, suppress its coherent evolution \cite{streed2006, raimond2012} or confine such coherences to Hilbert subspaces demarcated by measurement-induced boundaries \cite{facchi2002,schafer2014,signoles2014}. These phenomena are different manifestations of the Quantum Zeno Effect (QZE) \cite{misra1977,itano1990}. In addition to its foundational implications on the nature of quantum mechanics and the measurement process, the QZE has also garnered attention as a means of stabilizing fragile quantum states, studying emergent classicality in a quantum system due to measurement \cite{bhatt2000,jacobs2011,javanainen2013} and for controlling the thermodynamic properties of an isolated quantum system \cite{erez2008}.

In a broader context, a measurement can be regarded as a dynamically tunable interaction between a quantum system and a `bath' whose intrinsic, spatial and dynamical properties can be precisely engineered. As such, measurements can be used to guide or coax a quantum system into novel collective phases and non-equilibrium states that might otherwise be inaccessible through more conventional means of cooling or state preparation. While most measurement-induced control schemes have hitherto been demonstrated in the context of single, or weakly-interacting quantum entities, the extension of these concepts to the arena of strongly interacting and correlated many-body systems promises tantalizing prospects. 

Ultracold atomic and molecular gases in optical lattices have emerged as a clean, paradigmatic realization of correlated quantum many-body systems \cite{blochrmp2008}. The inherent control and tunability of various properties of these gases have allowed for a diverse range of studies focused on the realization of ultracold analogues of correlated electronic materials \cite{lewenstein2007}, studies of non-equilibrium dynamics of isolated quantum many-body systems \cite{polkovnikov2011}, and the creation of novel many-particle states of matter. These efforts have been further bolstered by novel developments of {\em in situ} lattice imaging techniques \cite{nelson2007,bakr2009,gemelke2009, wurtz2009, sherson2010}. To date, most of these demonstrations have relied on molasses cooling or absorption imaging, processes that invariably impart a large energy to each atom. Thus, while these techniques represent very valuable diagnostic tools, they are not currently amenable to measurement-based quantum control. 

In this work, we use a two-photon {\em in situ} lattice imaging technique to demonstrate the suppression and control of quantum tunneling in a lattice gas by the Quantum Zeno effect. In contrast to molasses-based lattice imaging schemes, our imaging technique extracts fluorescence from the lattice gas while retaining the atoms in the ground vibrational band of the lattice \cite{QNDimaging}. By extending this technique down to shallow lattice depths with correspondingly large tunneling rates, we show that the process of imaging, i.e. extracting spatial information from the lattice gas, has the concomitant effect of dramatically changing the tunneling dynamics. By taking advantage of the large dynamic range ($\sim \mathcal{O}(10^4)$) of measurement rates that are made available by our technique, we observe the continuous crossover of tunneling dynamics from the `weak measurement' regime, where the act of measurement exerts negligible backaction on the lattice gas, to the `strong measurement' or Zeno regime, where the act of measurement leads to a strong suppression of tunneling. 

\begin{figure}[t]
\includegraphics[width=3.5in]{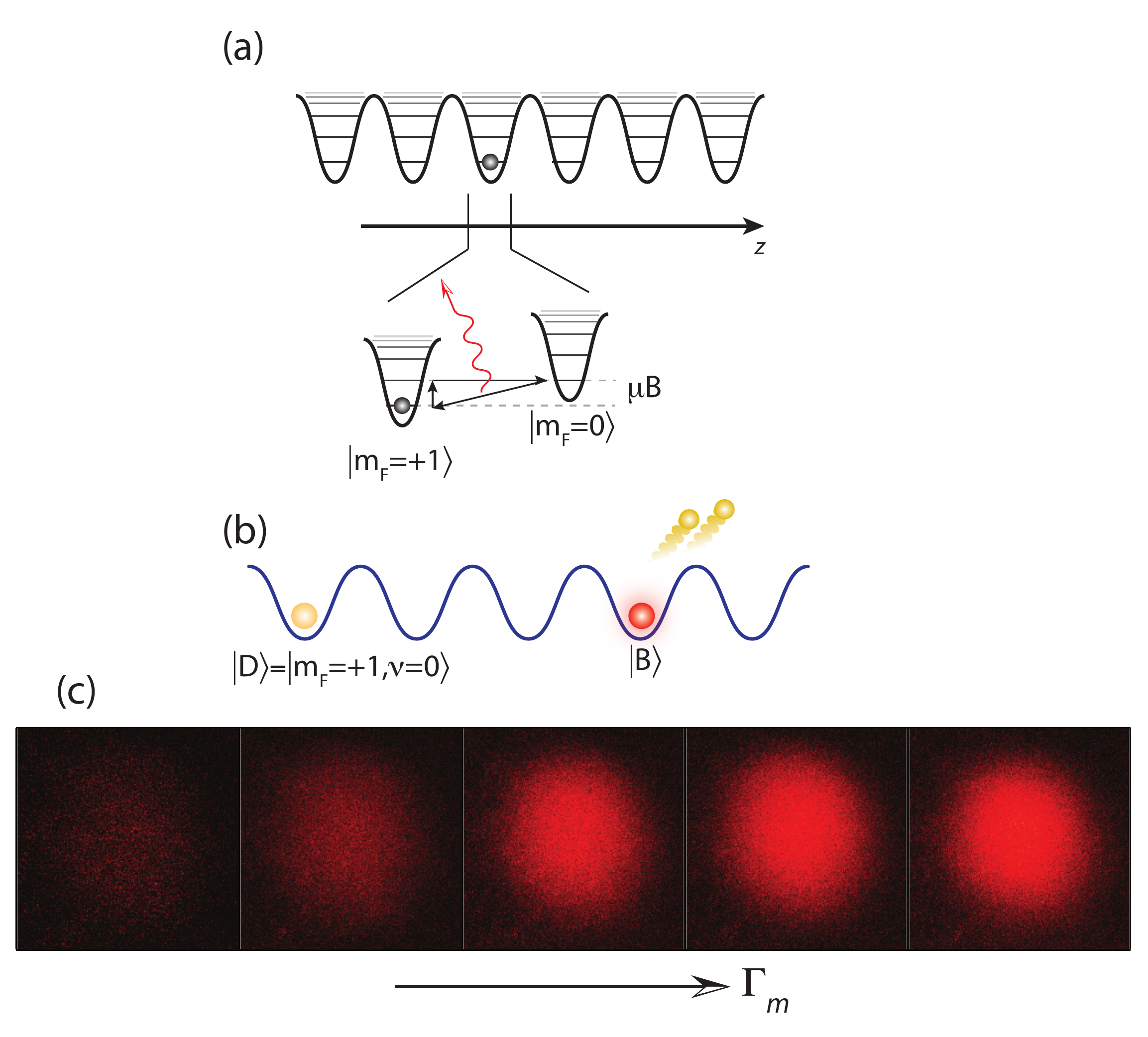}
\caption{(a) Lattice imaging scheme : An atom within a lattice site is cooled to the ground state $|D\rangle \equiv |F=1,m_F=+1; \nu = 0 \rangle$ via RSC. This state is nominally a `dark state', i.e. it does not emit fluorescence. An auxiliary `imaging' beam promotes the atom out of this state to a fluorescing state $|B\rangle$ which is subsequently cooled back to $|D\rangle$. Repeated cycles of this process extract fluorescence from the atom while continually restoring the atom to $|D \rangle$. (b) The imaging scheme thus allows us to distinguish between two possible states of the atom - a bright state $|B\rangle$ that can be imaged and a dark state $|D\rangle$ that cannot be imaged. (c) Fluorescence images of a lattice gas obtained at increasing levels of the measurement rate $\Gamma_m$. The field of view of each frame is 250 $\mu$m $\times$ 250 $\mu$m.}
\label{Fig:Fig1}
\end{figure}

The principle of the imaging scheme is depicted in Fig. 1. Raman sideband cooling (RSC) \cite{vuletic1998,hamann1998,han2000,kerman2000} is used to cool atoms within an optical lattice to the lowest vibrational band while simultaneously pumping them to the high field seeking spin state $|D\rangle \equiv |F=1,m_F=1; \nu=0\rangle$.  This state is decoupled from the light field and as such, does not emit fluorescence. As shown in \cite{QNDimaging}, fluorescence can be induced by controllably promoting the atoms to a bright state $|B\rangle$ and subsequently re-cooling them back to $|D\rangle$. 

The fluorescence emitted by the atoms can be captured by a detector and constitutes a position measurement of the emitting atom. While such position measurements nominally impart energy to the atom as a measurement backaction, the simultaneous use of RSC mitigates this increase in energy by cycling the atoms back to $|D \rangle$. Due to this cycling, fluorescence can be repeatedly extracted from the atomic distribution while leaving the atom in its original state. Thus, based on the spin- and vibrational state of the atoms, our imaging scheme distinguishes between a dark state $|D\rangle$ whose position cannot be measured, and a bright state $|B\rangle$ that emits fluorescence and hence, whose position can be measured. For such bright state atoms, we introduce a position measurement rate $\Gamma_m$. We define this to be the scattering rate of photons from the imaging beam \cite{scatnote}, and note that this underestimates the actual scattering rate of atoms since it neglects the spontaneous emissions during the subsequent re-cooling of atoms to $|D\rangle$. 

\begin{figure}[b]
\includegraphics[width=0.28\textwidth]{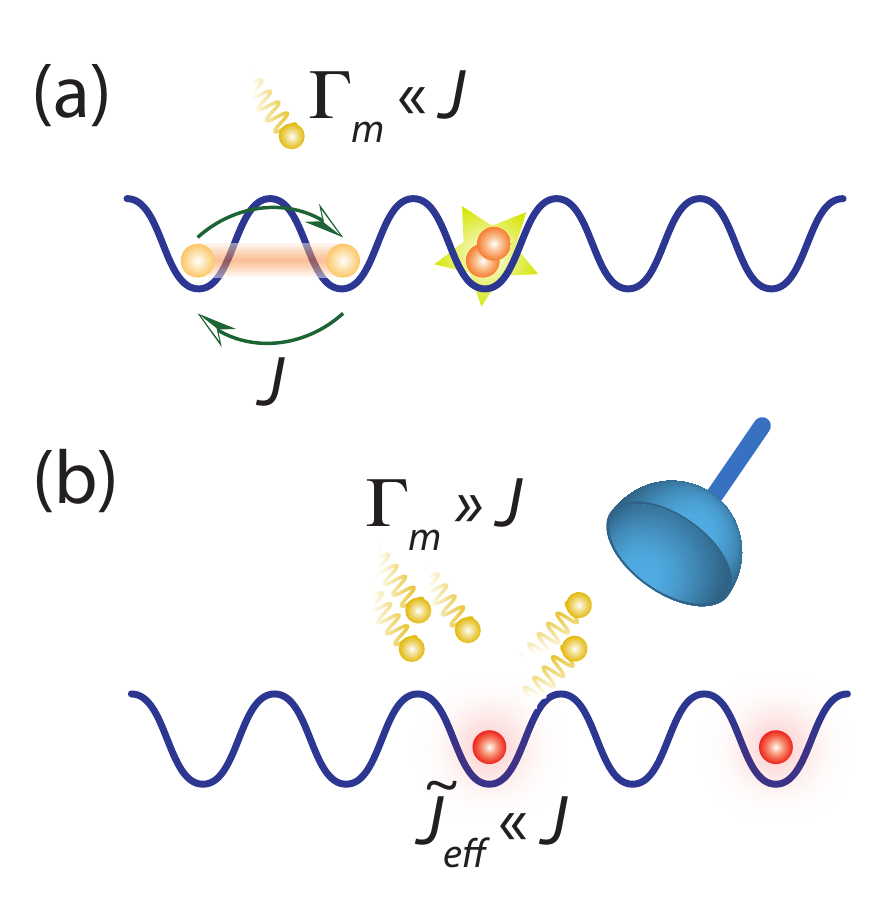}
\caption{Regimes of continuous position measurements of the lattice gas. (a) In the weak measurement regime, the rate of fluorescence ($\Gamma_m$) of an atom is much smaller than the coherent tunneling rate ($J$). Continuous position measurements thus have little influence on coherent tunneling dynamics and the photoassociation rate $\kappa$ is proportional to the bare lattice tunneling rate $J$. (b) In the strong measurement regime, the fluorescence rate is much larger than the tunneling rate. The frequent position measurements cause the atoms to be repeatedly projected back to the same lattice site, thereby strongly suppressing coherent tunneling and photoassociation.}
\label{fig:qze}
\end{figure}

In shallow lattices, the atomic distribution is not static. The atoms can coherently tunnel across sites at a rate $J$ that is exponentially dependent on the lattice parameter $s=U_0/E_r$ where $U_0$ is the depth of the lattice and $E_r$ is the recoil energy of the atom \cite{jaksch1998}. For the filling fractions used in this work ($f \sim 0.25$), such tunneling events frequently lead to multiply-occupied lattice sites and subsequent atom loss due to photoassociation. The photoassociation rate $\kappa$ (and two-body lifetime $\tau$) is given by the relation $\kappa = \tau^{-1} \approx 4 q J f$ where $q=6$ is the number of nearest-neighbors in our 3D lattice geometry. Thus, we use measurements of photoassociation within the lattice gas as a very sensitive probe of the tunneling rate. 

Quantum coherent tunneling of atoms within the lattice can be strongly influenced by continuous projective measurements of atomic position. Depending on the relative magnitudes of the tunneling rate $J$ and the measurement rate $\Gamma_m$, we can identify two distinct regimes. In the weak measurement limit $\Gamma_m \ll J$, the sporadic position measurements have negligible influence on tunneling, and the photoassociation rate $\kappa$ is independent of the measurement rate (Fig. 2(a)). In the strong measurement limit $\Gamma_m \gg J$, repeated position measurements continually project the atom into the same lattice site and strongly suppress coherent tunneling (Fig. 2(b)). In this regime, the coherent tunneling is supplanted by an incoherent `effective' tunneling rate given by $\tilde{J}_{eff} \sim J^2/\Gamma_m$ \cite{gagen1993, cirac1994}. In other words, the effective tunneling rate $\tilde{J}_{eff}$, and hence the photoassociation rate $\kappa$, monotonically decrease with increasing measurement rate. In essence, the act of observation `freezes' the lattice gas. This quantum phenomenon, which does not have a classical equivalent, is a manifestation of the Quantum Zeno effect. 

\begin{figure}[t]
\includegraphics[width=0.50\textwidth]{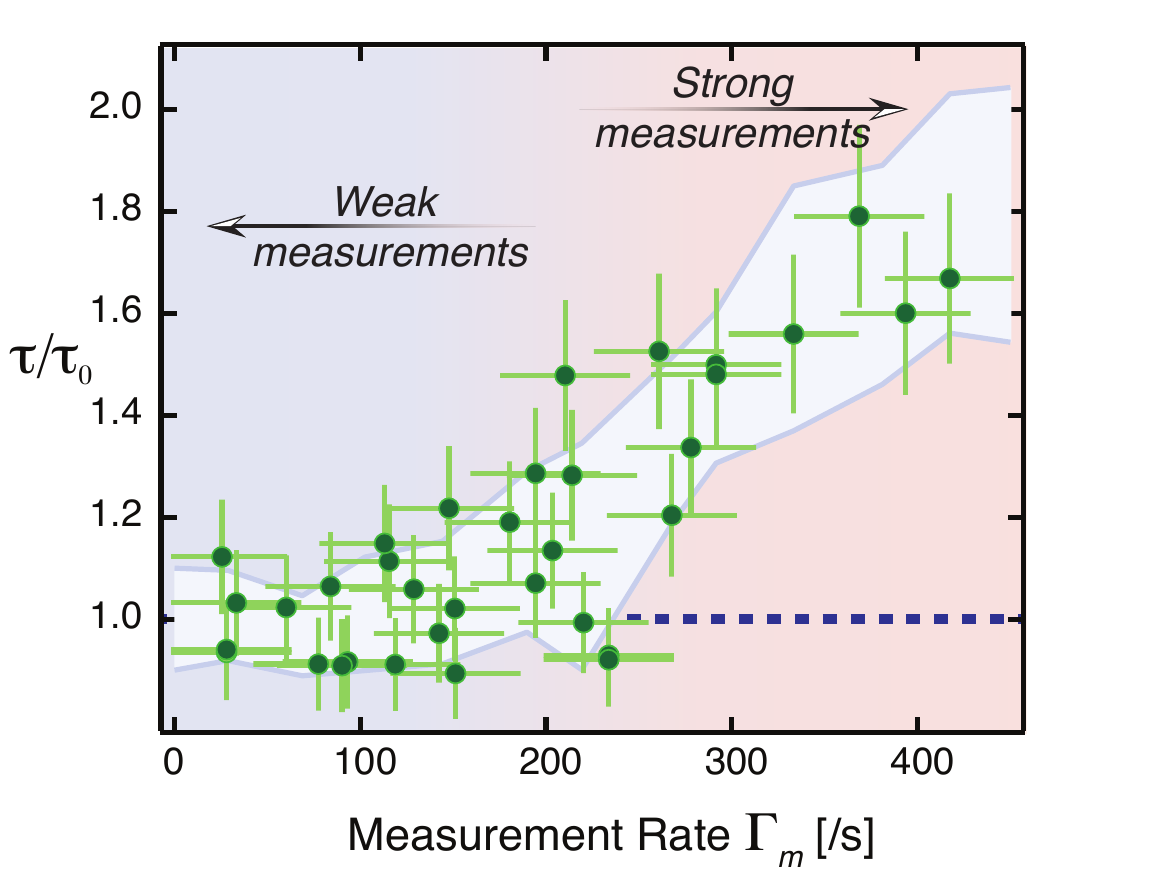}
\caption{Photoassociation measurements demonstrating the crossover from the weak measurement regime ($\Gamma_m \ll J$) to the strong measurement regime ($\Gamma_m \gg J$). In the former regime, the two-body lifetime of the lattice gas $\tau$ (inversely proportional to the photoassociation rate $\kappa$) is independent of the imaging rate, reflecting the negligible backaction induced by position measurements on the atoms. As the measurement rate becomes comparable to the coherent tunneling rate, the act of observation suppresses tunneling rates leading to an increase in the lifetime. These data were obtained at a lattice parameter $s = 8.5 (2.0)$ with $\tau_0 = 31(3)$ ms.}
\label{fig:qze}
\end{figure}

In our experiments, we prepare ultracold gases in the ground vibrational band of a 3D lattice (see Methods). In the absence of the imaging sequence, the lattice gas has a characteristic two-body lifetime $\tau_0$ that is dependent on the lattice depth (and corresponding `bare' tunneling rate) and residual light scattering due to Raman sideband cooling. In the simultaneous presence of sideband cooling, the lattice gas is subjected to either continuous or pulsed position measurements by the lattice imaging sequence at various measurement rates. This rate can be tuned over a large dynamic range ($\mathcal{O}(10^4)$) by varying the intensity of the optical field that induces fluorescence,  allowing us to probe both the weak and strong measurement limits as well as the crossover regime.

\begin{figure}[t]
\includegraphics[width=0.40\textwidth]{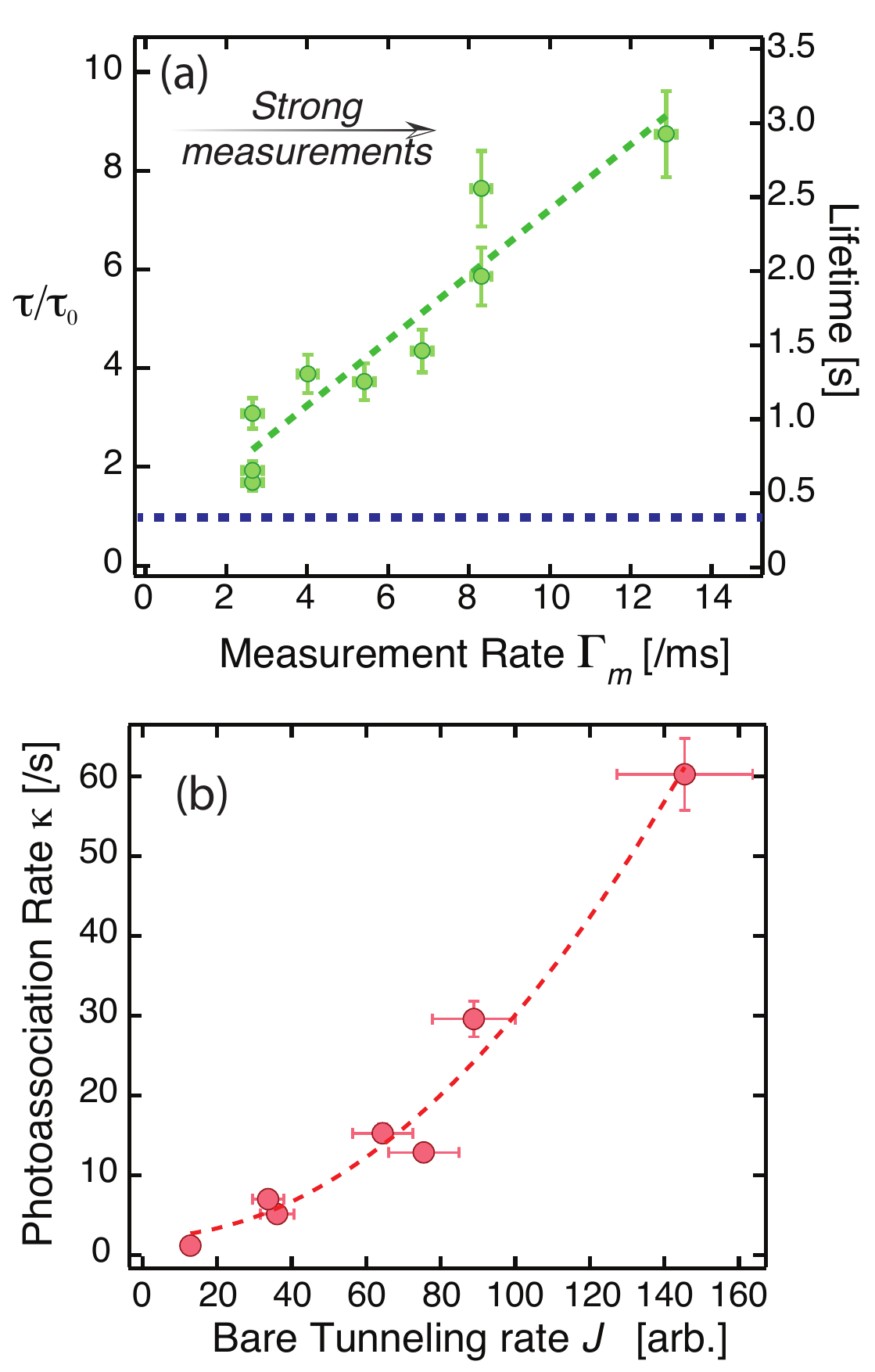}
\caption{In the strong measurement regime, the effective tunneling rate is given by $\tilde{J}_{eff} \sim J^2/\Gamma_m$. This leads to a two-body lifetime $\tau = \kappa^{-1}$ that linearly increases (as seen in (a)) with the measurement rate - a clear signature of the QZE. These data were obtained for $s=23(2)$.
(b) The quadratic scaling of the effective tunneling rate (and hence, the photoassociation rate $\kappa$) with the bare lattice tunneling rate is demonstrated by measurements of $\kappa$ for lattice gases confined in different lattice depths. These data were obtained by imaging the lattice gases at fixed measurement rate $\Gamma_m$. The dashed line shows a quadratic fit to the data.}
\label{fig:qze}
\end{figure}

At low rates of imaging, we observe that the two-body lifetime is unchanged by measurement, reflecting the negligible influence of position measurements on coherent tunneling. However, as the imaging rates increase, the two-body lifetime of the lattice gas is seen to grow (Fig. 3). This reflects the crossover from the weak measurement regime to the strong measurement regime where now, the measurement-induced localization of the atoms is the dominant influence on tunneling dynamics. This crossover regime offers a novel platform for quantitative studies of measurement-induced emergent classicality in a quantum system. 

As the measurement rate is made much larger than the coherent tunneling rate, the measured lattice lifetime grows in linear proportion to the measurement rate (Fig. 4(a)). This is a characteristic signature of the Quantum Zeno effect, i.e. increasing the rate of fluorescence {\em decreases} the photoassociation rate. As can be seen, the frequent projective measurements due to the imaging sequence can result in nearly a ten-fold suppression of coherent tunneling. 

Furthermore, the effective tunneling rate in the Zeno regime also exhibits a quadratic dependence on the bare tunneling rate. In order to demonstrate this, we confine lattice gases at varying depths while imaging the atoms at constant measurement rate. The measured photoassociation rates, shown in Fig. 4(b), exhibit a clear quadratic dependence with the bare tunneling rates estimated based on our calibration of the imposed lattice depth. This provides an additional corroboration for the QZE. 

At first glance, it would appear that an atom can be localized to a lattice site for arbitrary lengths of time for sufficiently large measurement rates. However, in general, the act of position measurement causes the atom's energy to increase linearly with time \cite{yanay2014}. In our scheme, this increase in energy is mitigated by the simultaneous use of sideband cooling at a rate $\Gamma_{RSC}$. For measurement rates that exceed this cooling rate, there is a significant probability of higher band occupation with correspondingly larger rates of tunneling. This causes a deviation from the linear growth of two-body lifetime with measurement rate (Fig. 5). Monte Carlo simulations of this competition between measurement-induced heating and Raman cooling are in good agreement with our observations. In the regime $\Gamma_m \gg \Gamma_{RSC}$, the measurement-induced heating dominates any cooling mechanism and the atom is completely delocalized due to rapid higher-band tunneling, leading to high rates of photoassociation (Fig. 5 (inset)). Based on these considerations, it is clear that the Zeno effect is most readily seen for the regime $J \ll \Gamma_m \ll \Gamma_{RSC}$. 

\begin{figure}[t]
\includegraphics[width=3.0in]{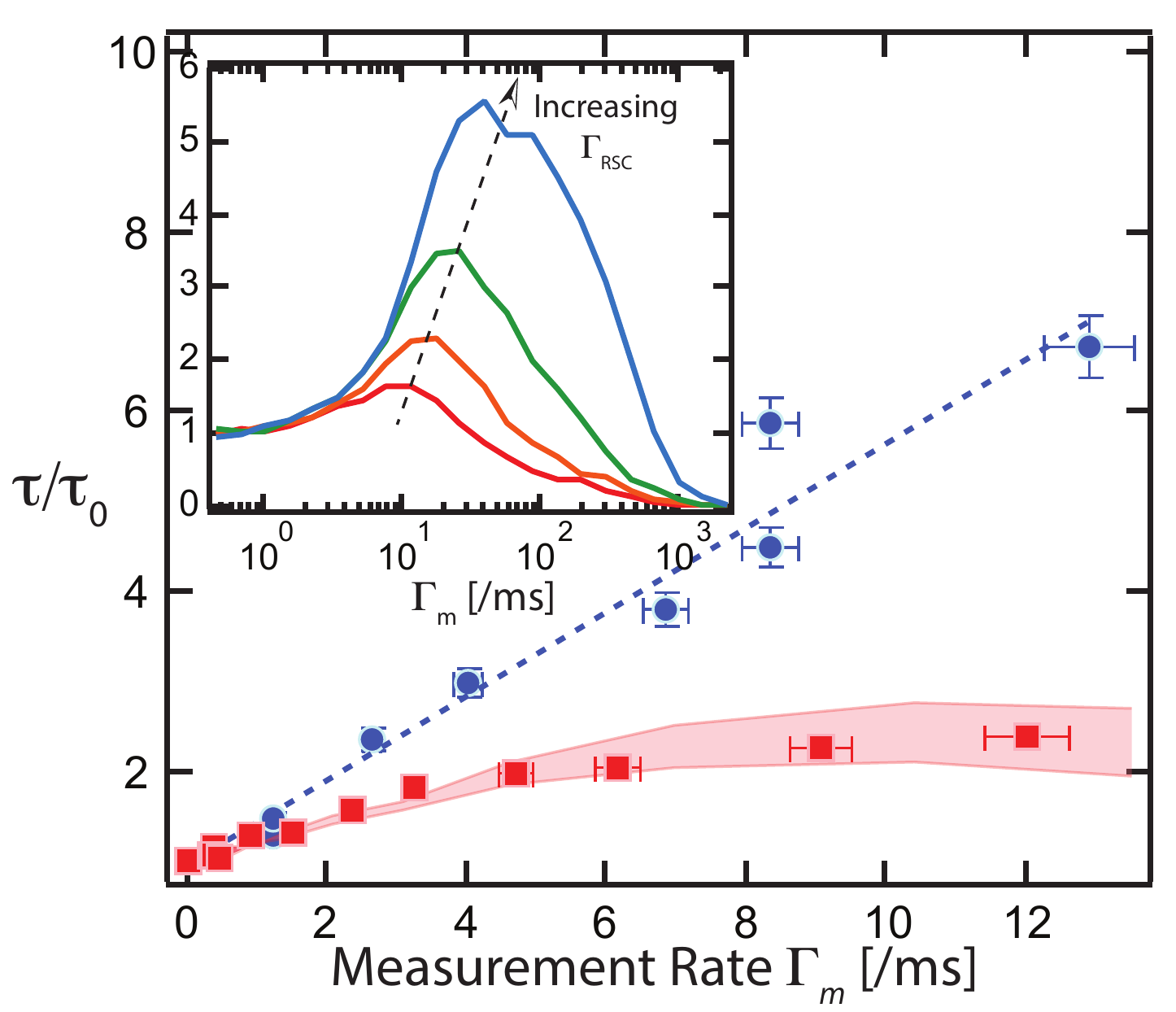}
\caption{For measurement rates $\Gamma_m$ that exceed the Raman cooling rate $\Gamma_{RSC}$, atoms are promoted to higher vibrational bands due to the measurement. The increased tunneling rates in these higher bands cause a deviation from the linear scaling of the lifetime $\tau$ with $\Gamma_m$. Due to the proportionate relation between the Raman cooling rate and the lattice depth, this deviation occurs more readily for atoms in shallow lattices. The data shown represent two-body lifetimes in the Zeno regime for lattice parameters $s=9.5(1.5), (\blacksquare)$, $s=21(2)$, ($\bullet$). The shaded region represents a Monte Carlo simulation of a kinetic model of the measurement process. Inset: Simulated two-body lifetimes {\em vs} measurement rate : the onset of higher-band tunneling occurs at larger $\Gamma_m$ for increasing Raman cooling rates (bottom to top).}
\label{Fig:nonlintau}
\end{figure}

In summary, we use an {\em in situ} lattice imaging technique to demonstrate the measurement-induced control of an ultracold lattice gas. 
By varying the rate of imaging, i.e. position measurements, in relation to the tunneling rate within the lattice, we show the smooth crossover from the weak measurement regime where the act of observation causes negligible backaction on the lattice gas, to the strong measurement or Zeno regime where measurement-induced localization causes a strong suppression of coherent tunneling. In contrast to previous demonstrations of the Zeno effect in a lattice gas due to intrinsic dissipative interactions within the lattice gas \cite{syassen2008,yan2013}, our demonstration relies on direct position measurements, i.e. imaging. The large dynamic range and quantum-limited tunability inherent to this imaging scheme thus enable new forms of measurement-induced control of a lattice gas by spatially and dynamically varying measurement landscapes. 

In addition to its foundational significance to the nature of quantum measurements, we also note the relevance of this study in the context of state preparation in ultracold many-body systems. While the isolation of such systems from the environment has notably allowed for the observation of various forms of long-lived mesoscopic quantum behavior, this decoupling also stymies the creation of low entropy states within experimentally viable timescales. In this regard, the identification of the measurement process as a `designer bath' might provide valuable perspectives on a truly bottom-up approach to quantum many-body systems with complete control over both intrinsic and external interactions. The techniques demonstrated here pave the way towards using quantum measurements for cooling, state preparation and spatially-resolved entropy segregation in a lattice gas. Further, they also augur intriguing prospects of realizing novel many-body interactions such as a measurement-induced dynamic coupling between the internal, motional and topological states of a quantum many-particle system. 

This work was supported by the ARO MURI on Non-equilibrium Many-body Dynamics (63834-PH-MUR), the DARPA QuASAR program through a grant from the ARO and the Cornell Center for Materials Research with funding from the NSF MRSEC program (DMR-1120296). We acknowledge valuable discussions with H. F. H. Cheung and I. S. Madjarov. M. V. acknowledges support from the Alfred P. Sloan Foundation. 

\bibliography{QZEbib,Imagingnotes}

\begin{thebibliography}{33}
\bibitem [{sca()}]{scatnote}%
  \BibitemOpen
  \href@noop {} {}\bibinfo {note} {The measurement rate is calibrated based on
  the measured power, beam profile and polarization of the imaging beam, and
  its orientation relative to the ambient magnetic field. At the lowest rates,
  allowing for various losses and stray light scattering, we estimate that this
  calibration is accurate to within a factor of two.}\BibitemShut {Stop}%
\end{thebibliography}

\section*{Methods}
\subsection*{Experimental sequence}
In our experiments, $^{87}$Rb atoms are confined in a 3D optical lattice that is typically detuned $2 \pi \times 160$ GHz from the $F=1 \rightarrow F'$ (D2) transition of $^{87}$Rb. The optical lattice provides both the confinement and the coherent two-photon coupling required for sideband cooling. In the absence of RSC, the measured heating rates are on the order of 5 nK/ms, about three orders of magnitude below the measured Raman cooling rates. By varying the loading conditions, the number of atoms confined in the lattice can be controlled from $10^4 - 10^8$. For the studies described here, atoms are first cooled to the ground vibrational band (average vibrational occupancy $\langle n_0 \rangle < 0.01$) in a deep lattice ($s \sim 50\pm1.5$) before the lattice is slowly ramped down to a final depth in the range $s = 6 - 20$. Depending on the final lattice depth, the ambient magnetic field and the optical pumping rates are adjusted continuously during the ramp so as to ensure high fidelities of ground state cooling. Typical filling fractions at the end of this sequence is around $f = 0.25(4)$, as estimated from measurements of atom number and cloud size.

The atomic ensemble is then subjected to either pulsed or continuous position measurements, i.e. imaging, as described in \cite{QNDimaging}, for variable periods of time. At the end of this imaging sequence, the atom number and temperature of the ensemble are measured using time-of-flight imaging and sideband spectroscopy. \\

\subsection*{Temperature and lattice depth calibration}
The average vibrational occupancy is probed through both time-of-flight measurements as well as sideband spectroscopy. For lattice depths $s > 10$, both these measurements coincide at the few percent level. For the lowest lattice depths ($s<10$), thermometry through measurements of sideband asymmetry was found to be less accurate as the spectral width of the sidebands was comparable to the lattice frequency. For such shallow lattices, we relied primarily on temperature measurements obtained by time-of-flight absorption imaging. 

The lattice depths were calibrated via sideband spectroscopy. These measurements were performed most accurately for the deeper lattices to obtain a calibration between the measured lattice frequencies and the optical power of the imposed lattice beams. This calibration was then extrapolated to the lowest optical powers to overcome the finite resolution of the sideband spectra for lattice depths $s<10$. 

\end{document}